\documentclass{aa}

\usepackage{graphicx}
\usepackage{txfonts}
\usepackage{mathtools}
\usepackage{bm}
\usepackage{natbib}
\bibpunct{(}{)}{;}{a}{}{,} 
\usepackage{hyperref}
\usepackage{float}

\begin{document}

   \title{
      Do tides play a role in the determination of the pre-stellar core mass function?
   }
   \titlerunning{Do tides play a role in the determination of the core mass function?}


   \author{Pierre Dumond \inst{1} \and Gilles Chabrier\inst{1,2} }

   \institute{CRAL, Ecole normale sup\'erieure de Lyon, Universit\'e de Lyon, UMR CNRS 5574, F-69364 Lyon Cedex 07, France\\
              \email{pierre.dumond@ens-lyon.fr}
         \and
             School of Physics, University of Exeter, Exeter, EX4 4QL, UK\\
             \email{chabrier@ens-lyon.fr} }

   \date{Received XXX; accepted XXX}

  \abstract{ 
   Recent studies have examined the role of tides in the star formation process. They suggest, notably, that the tides determine the characteristic mass of the stellar initial mass function (IMF) by preventing the collapse of density fluctuations that would become gravitationally unstable in the absence of the tidal field generated by a neighboring central mass. 
   However, most of these studies consider the tidal collapse condition as a 1D process or use a scalar virial condition and thus neglect the anisotropy of the tidal field and its compressive effects.
   In the present paper, we consider a turbulence-induced density perturbation formed in the envelope of a central core. This perturbation is subject to a tidal field generated by the central core. We study its evolution taking dynamical effects and the anisotropy of the tides into account.
   Based on the general tensorial virial equations, we determine a new collapse condition that takes these mechanisms into account. We identify two regimes: (i) a weak tidal regime in which the dynamics of the perturbation is only slightly modified by the action of the tides and (ii) a strong tidal regime in which the density threshold for collapse can potentially be increased due to the combined effects of the tides and the rotational support generated by the tidal synchronization of the perturbation with the orbital motion. In the case of a turbulence-induced density perturbation of mass $M_{\rm p}$ formed in the vicinity of a first Larson core, which is the case considered in some star formation scenarios, we show that the density threshold above which the perturbation collapses is increased only for low-mass perturbations ($M_{\rm p}\lesssim 2.7\,{\rm M}_\odot$) and only by  at most a factor of 1.5. We conclude that tides likely do not play a major role in the process of star formation or in the determination of the characteristic mass of the IMF. We propose an alternative explanation for the observed value of the characteristic mass of the IMF.
   }

   \keywords{Methods: analytical --- Stars: formation, luminosity function, mass function --- Turbulence
               }

   \maketitle
%
\nolinenumbers
\section{Introduction}

Gravity plays a major role in the star formation process. However, the way in which it affects the various stages of evolution of this process remains unclear. On one hand, it triggers the collapse of unstable perturbations, which leads to the formation of stars. On the other hand, it has been suggested that the long-range gravitational force exerted by an external body can disrupt such a perturbation through its tidal action \citep{Li_TidesCloudsControl2023}. 
Such a mechanism of tidal disruption has been widely studied in the case, for instance, of a star passing close to a supermassive black hole (see \citealt{Gezari_TidalDisruptionEvents2021} for a review) or of a molecular cloud located close to a black hole \citep{Tatematsu_DynamicsRotatingGaseous1990, Usami_TidalEffectsRotating1997, Chen_STABILITYGASCLOUDS2016a}, in particular in the context of the Central Molecular Zone \citep{Kruijssen_DynamicalEvolutionMolecular2019,Dale_DynamicalEvolutionMolecular2019a}. 

The effect of tides has recently been studied in more classical environments and has been invoked to explain the universality of the peak of the initial mass function \citep[IMF;][]{Lee_StellarMassSpectrum2018, Hennebelle_HowFirstHydrostatic2019, Colman_OriginPeakStellar2020}. In this picture, the tides prevent the formation of an unstable perturbation too close to an already formed Larson core. The material contained within this ``tidal radius'' is thus accreted by the central core, rather than forming a new core, resulting in a typical mass of about \(7-9 M_{\rm L}\), where \(M_{\rm L}\) denotes the mass of the first Larson core. The effect of tides on a density perturbation has also been studied in a more general context by \cite{Jog_JeansInstabilityCriterion2013} and \cite{Zavala-Molina_EffectTidalForces2023a}. However, these studies neglect an important feature of the tidal field, namely that it is necessarily anisotropic when it has an extensive component. As we will show, because of this lack of anisotropy, the collapse criterion derived in the presence of tides significantly overestimates the minimum density required for the perturbation to become gravitationally unstable. This issue has already been noted by \cite{Li_ModificationJeansCriterion2024}, who proposed an alternative way to derive the collapse criterion based on the Jeans analysis but taking the anisotropy into account.

To understand the role of tides in the dynamics of density perturbations, numerical simulations \citep{Ganguly_SILCCZoomDynamicBalance2024} have been used to investigate the 3D tidal field, by calculating the three eigenvalues of the external tidal tensor acting on a perturbation. They confirm that the tidal field is highly anisotropic, but the tidal energy is not strong enough compared to gravity to disrupt the structures. According to their study, the tides do not significantly modify the dynamics of unstable perturbations.

Estimates of the tidal strength acting on structures have also been made in several observational studies, based on either 2D maps \citep{Zhou_GasInflowsCloud2024} or 3D reconstructions of the density field \citep{Li_TidesCloudsControl2023}. Contrary to the study based on numerical simulation, both of these studies suggest that tides play an important role in the dynamics of the density perturbations. However, the first study relies purely on an energetic argument, namely on the fact that the turbulent energy is lower than the tidal energy. The second considers the amount of gas subject to extensive tides without analyzing the role of the latter on the formation of bound structures. Therefore, these studies cannot be considered robust analyses of the role of tides on structure formation.

In this paper we derive a new collapsing barrier condition, taking into account the dynamics of the expanding and contracting axes and the anisotropy of the tides{ based on the tidal equation derived in \cite{Chabrier_ConsistentExplanationUnusual2024} and using a fragmentation threshold based on the analysis of \cite{Inutsuka_SelfsimilarSolutionsStability1992}}. In Sect. \ref{Sec_Mathematical_model} we recall the usual mathematical description of the tidal field, and we derive the global energy budget of a perturbation in a tidal field in Sect. \ref{Sec_Global_energy_budget}. In Sect. \ref{Sec_Dynamical_evolution}
 we describe our dynamical model, which is based on the tensor virial equation, and compute a new collapsing barrier in the presence of tides. The role of the rotation of the perturbation is described in Sect. \ref{Sec_Tides_env}, and we discuss our results in Sect. \ref{Sec_Discussion}. The main notations used in the article are presented in Table \ref{tab_notation}.

\begin{table*}
\caption{Main notations used in the article.}
\label{tab_notation}
\begin{center}
\begin{tabular}{cc}
   \hline Notation & Description \\
    \hline G & Gravitational constant \\
    $\vec{g}$ & Gravitational field \\
    $\phi$ & Gravitational potential \\
    $\vec{g}_{\rm T}$ & Tidal field \\
    $c_{\rm s}$ & Sound speed \\
    \hline $M_{\rm c}$ & Mass of the core of the central object \\
   {$\rho_{\rm e}(r)\propto r^{-\gamma}$} & Envelope density {profile} of the central object \\
    $\gamma$ & Index of the density profile of the envelope of the central object \\
    $M_{\rm c+e}$ & Total mass of the central object (core + envelope) \\
    \hline $M_{\rm p}$ & Mass of a density perturbation formed in the vicinity of an already formed core (the central object)  \\
    $R$ & Radius of the spherical perturbation when it forms\\
    $a, b, c$ & Semi axes of the perturbation deformed by the tidal field \\
    $r_0$ & Distance between the barycenters of the perturbation and the central object \\
    $\bm{\Omega}$ & Rotation vector of the perturbation around the central object \\
    \hline $\mathbb{T}$ & Tidal tensor \\
    $\mathbb{T}_{\rm e}$ & Contribution to the tidal tensor from the envelope of the central object\\
    $\mathbb{T}_{\rm c}$ & Contribution to the tidal tensor from the core of the central object \\
    $\mathcal{T}$ & Virial Tidal tensor \\
    $\mathcal{I}$ & Virial Inertia tensor \\
    $\mathcal{K}$ & Virial Kinetic energy tensor \\
    $\mathcal{W}$ & Virial Gravitational energy tensor \\
    $\mathcal{R}$ & Rotation support virial tensor \\
   \hline
\end{tabular}
\end{center}
\end{table*}

\section{Mathematical model of the tides}
\label{Sec_Mathematical_model}

The physical origin of tides is the force resulting from the difference between the gravitational force acting at the center of mass of a structure and the one acting at another point within it. For a density perturbation characterized by its center of mass at position \(\vec{r}_0\) in a gravitational field \(\vec{g}(\vec{r})\) generated by an external mass distribution, the central object, the tidal field (\(\vec{g}_{\rm T}\)) associated with this gravitational field is defined as
\begin{equation}
   \label{eq_tidal_field_exact}
   \vec{g}_{\rm T}(\vec{r}_1) = \vec{g}(\vec{r}_1)-\vec{g}(\vec{r}_0),
\end{equation}
where \(\vec{r}_1=\vec{r}_0+\delta\vec{r}\) is the position of a given point inside the perturbation. 
The effect of this tidal field is to deform the structure by either compressing or disrupting its axes. 

In the usual cases of interactions between structures in the interstellar medium (ISM), we assumed a density perturbation whose size is much smaller than its distance to the object generating the gravitational field (i.e., \(|\delta\vec{r}| \ll |\vec{r}_0|\)). The range of validity of this limit is discussed in Appendix \ref{App_tidal_approx}. In this case, the tidal field can be written as 
\begin{equation}
   \label{eq_tidal_field_tensor}
   g_T^i(\delta\vec{r})=-\delta r_k\frac{\partial^2 \phi}{\partial x_k \partial x_i}(\vec{r}_0),
\end{equation}
where we introduce the gravitational potential \(\phi\).  The tidal tensor is usually defined as \(\mathbb{T}_{ij} = -\frac{\partial \phi}{\partial x_j \partial x_i}\). The tides are fully  compressive if the 3 eigenvalues of the tensor, along the three orthogonal directions, are positive, while they are extensive if one eigenvalue is negative.

In the ISM, the gravitational field is very complicated, as many massive structures (bound cores, clumps, density perturbations of all kinds) interact with each other. Following many authors \citep{Lee_StellarMassSpectrum2018,Colman_OriginPeakStellar2020,Zavala-Molina_EffectTidalForces2023a}, we considered only the interaction between two objects, namely a density perturbation of mass $M_{\rm p}$ and typical radius $R$ subject to the tides generated by a nearby central object. This simplification is justified by the rapid decay of the \(1/r^2\) gravitational force implying that the dynamics is dominated by the interaction between the closest objects. 
The central object consists of a core of mass $M_{\rm c}$ surrounded by an envelope characterized by a density profile \(\rho_{\rm e}\propto r^{-\gamma}\). This global structure (core + envelope) will be denoted as the central object of mass $M_{\rm c+e}$ all along this study.
In the case of pre-stellar dense cores, \(\gamma\) is usually observed to be close to 2 \citep{Roy_ReconstructingDensityTemperature2014,Pineda_BubblesFilamentsCores2022}, which corresponds to the gravitational collapse profile of a sphere \citep{Larson_NumericalCalculationsDynamics1969}. We use this value in the following. 

\begin{figure}
   \centering
   \includegraphics[scale=0.6]{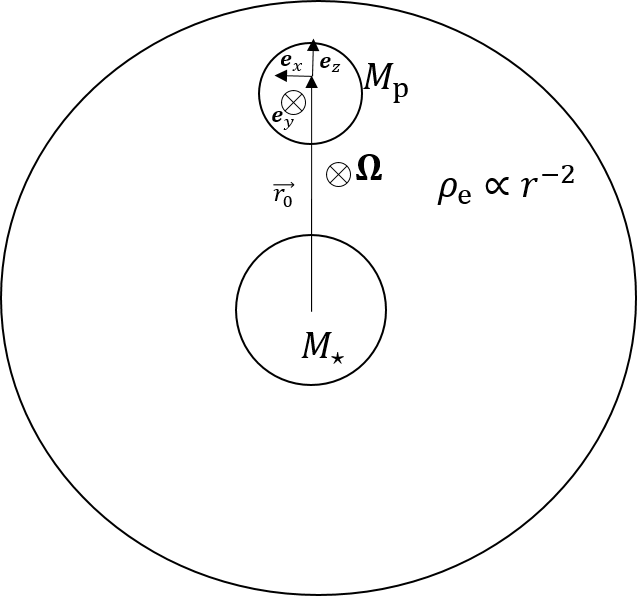}
   \caption{Schema of the physical situation studied in this paper. When it is discussed in Sect. \ref{Sec_Tides_env}, the rotation is oriented parallel to \(\vec{e}_y\).}
   \label{Fig_Schema}
\end{figure}

The derivation of the components of the tidal tensor in a Cartesian \( (\vec{e}_x, \vec{e}_y, \vec{e}_z) \) frame such that \(\vec{e}_z \parallel \vec{r}_0\) can be found in \citep{Colman_OriginPeakStellar2020,Zavala-Molina_EffectTidalForces2023a}. The situation is summarized in Fig. \ref{Fig_Schema}. The core contribution can be written as
\begin{equation}
   \mathbb{T}_{\rm c}=\frac{G M_{\rm c}}{r_0^3}\left[\begin{array}{ccc}
      -1 & 0 & 0 \\
       0 & -1 & 0 \\
       0 & 0 & 2
      \end{array}\right],
\end{equation}
where  \(r_0\) is the distance between the barycenters of the perturbation and the central object. The contribution from the envelope is
\begin{equation}
   \mathbb{T}_{\rm e}=\frac{4 \pi}{3-\gamma} G \rho_{\rm e}(r_0)\left[\begin{array}{ccc}
      -1 & 0 & 0 \\
      0 & -1 & 0 \\
      0 & 0 & (\gamma-1)
      \end{array}\right].
\end{equation}
The total tensor is  \(\mathbb{T} = \mathbb{T}_{\rm e} + \mathbb{T}_{\rm c}\).

In these two  tensors, the off-diagonal components are not strictly zero in general. They have high order terms, depending in particular on the ratios \(\delta\vec{r}_x/r_0\) and \(\delta\vec{r}_y/r_0\). Besides the fact that these terms can be neglected because we have assumed that \(|\delta\vec{r}| \ll |\vec{r}_0|\) (see above), this tensor is evaluated in \(\vec{r}_0 = (0, 0, r_0)\) after a Taylor expansion. In this case, the off-diagonal terms are exactly zero. 

The global evolution of the  structure can be computed thanks to the Viral Theorem (see, e.g., \cite{Chandrasekhar_ProblemsGravitationalStability1953,Lebovitz_VirialTensorIts1961, Lequeux_InterstellarMedium2005}). The tensor, \(\mathcal{T}_{ij}\), involved in the virial equation can be calculated as
\begin{align}
   \mathcal{T}_{ij} &= \int_{V}\rho(\delta\vec{r}) \mathbb{T}_{ik}(\vec{r}_0) \delta r_k \delta r_j {\rm d}V.
\end{align}
In the same frame as before, assuming that the perturbation is a homogeneous ellipsoid with semi axes \(a, b\) and \(c\), the tensor is expressed in Eq. \ref{eq_Virial_tidal_tensor}.
\begin{table*}
\begin{equation}
   \mathcal{T} = \frac{M_{\rm p}}{5}G\left[\begin{array}{ccc}
      -a^2\left(\frac{M_{\rm c}}{r_0^3}+\frac{4 \pi}{3-\gamma} \rho_{\rm e}(r_0)\right) & 0 & 0 \\
      0 & -b^2\left(\frac{M_{\rm c}}{r_0^3}+\frac{4 \pi}{3-\gamma} \rho_{\rm e}(r_0)\right)  & 0 \\
      0 & 0 & c^2\left(\frac{2M_{\rm c}}{r_0^3}+\frac{4 \pi(\gamma-1)}{3-\gamma} \rho_{\rm e}(r_0)\right)
      \end{array}\right].\label{eq_Virial_tidal_tensor}
\end{equation}
\end{table*}

The sign of each component determines the evolution of the corresponding axis. A positive component corresponds to the expansion of the axis, while a negative component indicates its contraction. The axes \(a\) and \(b\) are compressed by the action of the tides, while \(c\) (the axis oriented toward the central object generating the tides) can be either compressed or dilated depending on the slope of the profile and the mass of the central object. 
In the next section we give an indication of the global evolution of the structure from these three components. 

\section{The global energy budget from the tides}
\label{Sec_Global_energy_budget}

Before considering the evolution of the perturbation based on the tensorial Virial Theorem, taking into account the anisotropy of the tidal field, we first discuss the trace of the tensor \(\mathcal{T}\), which plays a major role in a stability analysis based on the scalar Virial Theorem (e.g., \cite{Lequeux_InterstellarMedium2005}). Indeed, a collapse criterion used in many star formation theories \citep{Hennebelle_AnalyticalTheoryInitial2008, Lee_StellarMassSpectrum2018} is derived from this theorem: a structure collapses if \(\partial^2\mathrm{Tr}(\mathcal{I})/\partial t^2<0\), where \(\mathcal{I}\) is the Virial inertia tensor. Knowing the initial properties of a newly formed perturbation, this criterion gives a first indication of its evolution. 
A positive trace of \(\mathcal{T}\) means that the tides favor the global disruption of the fluctuation (i.e., a decrease in its density), while a negative trace favors its global collapse. For a spherical fluctuation of radius \(R\), from Eq. (\ref{eq_Virial_tidal_tensor}) we have
\begin{equation}
   \text{Tr}(\mathcal{T}) = -\frac{M_{\rm p}}{5}G R^2 4\pi\rho_{\rm e}(r_0)<0.
\end{equation}In that case, regardless of the mass of the central object and the density profile, the tides will always favor the collapse of the fluctuation (i.e., its density will increase). More precisely, the fact that the trace of the tensor \(\mathcal{T}\) is not zero is related to the fact that the perturbation is "superimposed" on the envelope. This mass of the envelope engulfing the perturbation will favor the gravitational collapse. We note that the contribution of the central { mass} disappears in the global evolution of the perturbation. This is due to the fact that the trace of \(\mathbb{T}_{\rm c}\) is 0 while the envelope always favors global collapse, since the trace of \(\mathbb{T}_{\rm e}\) is always negative. 

However, this collapse criterion has a major shortcoming, since it does not take into account the deformation of the initial spherical perturbation due to the anisotropy of the tides. In the next section we address this problem with the tensorial form of the virial theorem. 

\section{The dynamical evolution of a static ellipsoid in a tidal field}
\label{Sec_Dynamical_evolution}

Most of the works investigating the role of tides in star formation models only consider the tidal acceleration of a central core and its envelope on a nearby perturbation. As will be examined in \S6.2, this situation is not physical: in that case, the perturbation would always collapse onto the central core. A more complete picture of the dynamics of the perturbation in the vicinity of a core should be considered, namely the addition of a rotation that supports the perturbation against its free fall onto the central core. This will be done in Sect. \ref{Sec_Tides_env}. Before considering this complete situation, we first consider the effect of the tides without taking into account the phenomena that prevent the collapse of the perturbation into the central core, in order to compare the results with what has been done in the literature.

A "static" (i.e., not orbiting) ellipsoid subject to an external tidal force corresponds to the seminal Jeans ellipsoid, and its equilibrium figure has been extensively studied in the incompressible case \citep{Chandrasekhar_EquilibriumStabilityJeans1963}. Here, we studied its dynamics in the compressible case according to the viral tensor equations. 

\subsection{The tensorial virial theorem}
\label{Sec_tensorial_virial_th}
The tensorial virial  theorem allows us to compute the evolution of the three diagonal components of the inertia tensor, \(\mathcal{I}\):
\begin{equation}
   \label{eq_Viral_eq}
   \frac{1}{2}\frac{\partial^2 \mathcal{I}_{ii}}{\partial t^2}= 2\mathcal{K}_{ii}+\mathcal{W}_{ii} +\int P{\rm d}V + \mathcal{T}_{ii}
,\end{equation} 
where \(\mathcal{I}_{ij}=\int \rho \vec{r}_i\cdot\vec{r}_j dV\) is the inertia tensor, \(2\mathcal{K}_{ij}=\int \rho\vec{v}_i\cdot\vec{v}_j dV\) the kinetic energy tensor and \(\mathcal{W}_{ij}=\int\rho r_i\frac{\partial \phi}{\partial_j}dV\) the potential energy tensor.
In the following, since we are focusing on the effect of the tides, we neglect the role of the surface term (external pressure) as it is independent of the tides. We also neglect the effect of any external accretion flow or magnetic field. 
From \cite{Chandrasekhar_EllipsoidalFiguresEquilibrium1969}, the potential energy tensor for a homogeneous ellipsoid of mean density $\rho$ is given by
\begin{equation}
   \mathcal{W}_{ii}=-\frac{2}{5}\pi G\rho M_{\rm p} \alpha_i c^2,
\end{equation}
with
\begin{equation}
   \alpha_i=\int_0^{\infty}\frac{du}{\sqrt{(1+u)(1+u\left(\frac{c}{a}\right)^2)(1+u\left(\frac{c}{b}\right)^2)}\left(1+u\left(\frac{c}{a_i}\right)^2\right)}.
\end{equation}
We note that \((a_1, a_2, a_3) = (a, b, c)\) are the semi-axes of the ellipsoids. The inertia  tensor of a homogeneous ellipsoid is
\begin{equation}
   I_{ii}=\frac{M_{\rm p}}{5}a_i^2.
\end{equation}
The kinetic energy tensor is assumed to be isotropic. It entails the turbulent velocity dispersion:
\begin{equation}
   2\mathcal{K}_{ii} = \frac{1}{3}M_{\rm p} v_{\rm RMS}^2.
\end{equation}
The pressure term describes  the pressure support:
\begin{equation}
   \int P{\rm d}V = M_{\rm p} c_{\rm s}^2,
\end{equation}
where we assume that the gas is isothermal. 

Denoting \(\tilde{x}\) the quantities normalized to the Jeans length, \(\lambda_{\rm J}\simeq c_{\rm s}/\sqrt{\bar{\rho}G}\), Jeans mass, \(M_J=4\pi/3\bar{\rho}\lambda_{\rm J}^3\), and mean free-fall time \(\tau_{\rm ff}^0\simeq 1/\sqrt{G\bar{\rho}}\),  where $\bar{\rho}$ is the mean density of the surrounding gas, one gets after calculations (see, e.g., \citealt{Chabrier_ConsistentExplanationUnusual2024})\begin{align}
   \label{eq_sys_axis_evol}
   \frac{1}{2}\frac{\partial^2 \tilde{a}^2}{\partial\tilde{t}^2} &=5(1+\mathcal{M}_\star^2 \tilde{a}^{2\eta})-2\pi\alpha_1 \tilde{c}^2\frac{\rho}{\bar{\rho}}-\tilde{a}^2(\mu_{\rm c}+\mu_{\rm e})  ,  \\
   \frac{1}{2}\frac{\partial^2 \tilde{c}^2}{\partial\tilde{t}^2} &=5(1+\mathcal{M}_\star^2 \tilde{a}^{2\eta})-2\pi\alpha_3 \tilde{c}^2\frac{\rho}{\bar{\rho}}+\tilde{c}^2\left(2\mu_{\rm c}+(\gamma-1)\mu_{\rm e}\right). \nonumber
\end{align}
Because the system is azimuthally symmetric, the axes \(a\) and \(b\) are equal during the evolution. 
We introduced the quantity \(\mathcal{M}_\star^2=\frac{V_0^2}{3c_{\rm s}^2}\left(\frac{\lambda_{\rm J}}{1\text{pc}}\right)^{2\eta}\), which is the Mach number at the Jeans length scale \citep{Hennebelle_AnalyticalTheoryInitial2008}, with the Larson velocity relation \(v_{\rm RMS}(R)=V_0 \left(\frac{R}{1\text{pc}}\right)^{\eta}\), where $R$ denotes the radius of the perturbation, with \(\eta=0.5\) { and \(V_0=1.6\, \text{km/s}\) is the 3D velocity normalization for parsec size structures \citep{Hennebelle_TurbulentMolecularClouds2012}}. The tidal coefficients related to the central core, \(\mu_{\rm c}\), and to the envelope, \(\mu_{\rm e}\), are given by
\begin{align}
   \mu_{\rm c} &= \frac{M_{\rm c}}{\bar{\rho} r_0^3}, \\
   \mu_{\rm e} &= \frac{4\pi}{3-\gamma}\frac{\rho_{\rm e}(r_0)}{\bar{\rho}}.
\end{align}
From the value of these coefficients we can distinguish two regimes: a strong tidal regime and a weak tidal regime. Indeed, the tides will be dominant (strong tidal regime) if the dynamical timescale, \(\tau_{\rm dyn}=\sqrt{\frac{r_0^3}{GM_{\rm c+e}}}\), is smaller than the collapsing timescale of the perturbation, \(\tau_{\rm ff}=\frac{1}{\sqrt{G\rho}}\). As mentioned previously, \(M_{\rm c+e}\) is the mass of the central core plus the envelope contained within a radius \(r_0\), while $\rho$ denotes the density of the perturbation.
The dynamical timescale characterizes the time it takes for the tides to significantly alter the dynamics of the perturbation. These two regimes can be thus summarized as follows: 
\begin{align}
   \label{eq_def_strong}
  &\text{Strong tidal regime}: \tau_{\rm ff}>\tau_{\rm dyn} \Leftrightarrow \tilde{\rho}= \frac{\rho}{\bar{\rho}}<\mu_{\rm c}+\mu_{\rm e} \\
  \label{eq_def_weak}
  &\text{Weak tidal regime}: \tau_{\rm ff}<\tau_{\rm dyn} \Leftrightarrow \tilde{\rho}= \frac{\rho}{\bar{\rho}}>\mu_{\rm c}+\mu_{\rm e}.
\end{align}
In the strong tidal regime, the tides are expected to play a significant role in the dynamics of the perturbation, while in the weak one they are not.

{ In all what follows, we consider a fiducial star forming cloud of size \(L_{\rm i}=10\,\text{pc}\) in which turbulence is injected at the size of the cloud. The gas filling the medium is assumed to be mostly dihydrogen, $H_2$, characterized by a mean molecular weight \(\mu=2.3\). According to the usual Larson relation linking  the size of the cloud to its mean density, the latter is \(\bar{\rho}=7\times 10^2 \text{cm}^{-3}\). This leads to \(\mathcal{M}_\star = 2.7\). The typical temperature of the medium is taken to be 10 K, leading to a typical sound speed \(c_{\rm s}=\)0.2 km/s assuming a thermodynamical coefficient of 7/5 for diatomic gas. These conditions are representative of typical Milky way star forming conditions given by \cite{Chabrier_VariationsStellarInitial2014}}.

\subsection{The evolution of the three axes}

As discussed in Sect. \ref{Sec_Global_energy_budget}, the tides tend globally to favor the collapse of an initially spherical structure since the trace of the virial tidal tensor \(\mathcal{T}\) is negative.  However, as the structure collapses, it will be deformed by the anisotropic effect of the tides. The axis \(c\) is elongated by the tidal field, while the  two other axes contract under its action. This deformation tends to stabilize the structure, which could eventually stop to collapse: at constant density and mass, an elongated ellipsoid is more stable than a spherical one. However, at least in the first stages of the evolution, the contraction of the axes \(a\) and \(b\) is faster than the expansion of the axis \(c\), which leads to an increase in the density of the perturbation. This is again closely related to the fact that \(\text{Tr}(\mathcal{T})\leq0\): the density of the structure increases at the very beginning of its evolution. If the density increase is large enough, the axis \(c\) will rapidly stop expanding and begin to collapse as well. 
The final state of a perturbation (collapsed or disrupted) is thus determined by a competition between the increase in density on one hand and its deformation by the tides on the other hand.
 
If the tidal field is fully compressive, which is possible if \(\gamma<1\), then each axis is compressed. In this case, the qualitative effect of the tides is isotropic. The calculation of a corrected Jeans length as done by \cite{Jog_JeansInstabilityCriterion2013} and \cite{Zavala-Molina_EffectTidalForces2023a} is then qualitatively relevant: the Jeans length will decrease because the structure will be more unstable.
Conversely, if the tidal forces are not fully compressive, the analysis performed by these authors is irrelevant because it does not capture the anisotropy of the process. The instability criterion cannot be reduced to a single threshold condition of one of the axes. This problem has already been pointed out by \cite{Li_ModificationJeansCriterion2024}. He proposed a way to account for anisotropy in the Jeans analysis but did not consider the dynamics of the process. In the present study, we propose another, more accurate way to calculate the barrier, one that takes the dynamics of the ellipsoid evolution under the action of the tides  into account.

By numerically solving the system of equations (\ref{eq_sys_axis_evol}) with a Runge-Kutta solver of order 2-3, we obtain the three possible evolutions of a perturbation in an anisotropic tidal field. At the end of the integration, all three axes have collapsed, only two have collapsed, or none has collapsed. If the perturbation is partially or globally collapsing, the integration is stopped when its density reaches the adiabatic density, \({\bar n}_{\rm ad} = 10^{11}\)cm\(^{-3}\). We note that we do not pretend to precisely calculate the dynamics of the collapse of the density perturbation. Indeed, when the density becomes very large and the collapse very fast, additional mechanisms may play a significant role in the dynamics, such as the adiabatic heating described by \cite{Robertson_ADIABATICHEATINGCONTRACTING2012}, and slow down the collapse. In addition, as discussed in the next section, the rapid collapse of two of the three axes can cause the aspect ratio to become very large. At this stage, fragmentation of the initial perturbation into several substructures is likely to occur. After fragmentation, our model no longer captures the evolution of the substructures.
Here, we only focus on the initial stage of the collapse of a perturbation under the action of a tidal field, and show that it affects the dynamics of the structure in a nontrivial way because of the anisotropy of the field.

\begin{figure}
   \centering
   \includegraphics[width=\columnwidth]{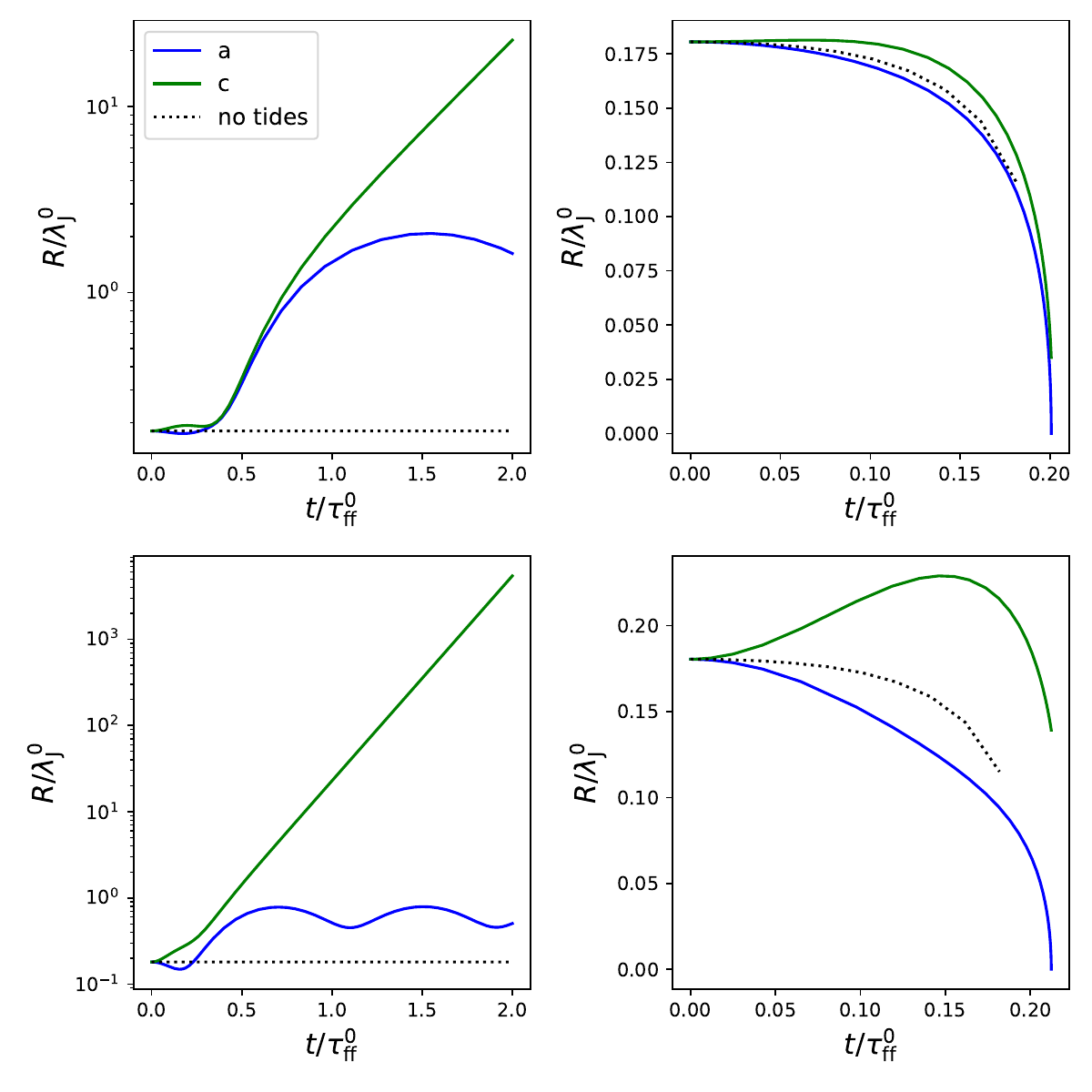}
   \caption{Evolution of two axes of the ellipsoid for different tidal field strengths and perturbation densities. The tidal field strength is characterized by \(\mu_{\rm c} = 5\) (top) and \(\mu_{\rm c} = 30\) (bottom). On the left, the initial density of the structure is equal to the critical density, \(\rho_{\rm HC}\), computed by \cite{Hennebelle_AnalyticalTheoryInitial2008}. Thus, the axis does not evolve in the absence of an external tidal field (dotted black line). On the right, the density is given as \(\rho = 0.9\rho_{\rm HC} + 0.1\rho_c\). Depending on the strength of the tidal field, the three axes may or may not collapse: if the tidal field is not too strong, the \(c\) axis collapses after a period of expansion at the beginning of the evolution.}
   \label{Fig_Evol_axis}
\end{figure}

In Fig. \ref{Fig_Evol_axis} we plot the evolution of the axes \(a\) and \(c\) of the ellipsoid for two tidal field strengths \(\mu_{\rm c} = 5\) (top) and \(\mu_{\rm c} = 30\) (bottom) and two densities of the perturbation normalized to the  mean density,  $\tilde{\rho}=\rho/{\bar \rho}$. In these plots, we set \(\mu_{\rm e}\) to 0. On the left, the density is equal to the critical density calculated by \cite{Hennebelle_AnalyticalTheoryInitial2008}:
\begin{equation}
   \label{eq_HC_collapse}
   \tilde{\rho}=\tilde{\rho}_{\rm HC} = \frac{15}{4\pi}\frac{1+\mathcal{M}_\star^2 \tilde{R}^{2\eta}}{\tilde{R}^2}.
\end{equation}
On the right, the density is a combination of \(\tilde{\rho}_{\rm HC}\) and \(\tilde{\rho}_{c}\), which is the critical density along the \(c\) axis that can be calculated from Eq. (\ref{eq_sys_axis_evol}) by solving \(\partial^2 c^2/\partial t^2=0\):
\begin{equation}
   \tilde{\rho}_{\rm c} = \frac{15}{4\pi}\frac{1+\mathcal{M}_\star^2 \tilde{a}^{2\eta}}{\tilde{c}^2} + \frac{3}{4\pi}\left(2\mu_{\rm c}+(\gamma-1)\mu_{\rm e}\right).
\end{equation}
We have chosen \(\tilde{\rho} = 0.9\tilde{\rho}_{\rm HC} + 0.1\tilde{\rho}_{\rm c}\). We note that \(\tilde{\rho}_{\rm c} \) is the density threshold chosen by \cite{Colman_OriginPeakStellar2020} in their analysis. Physically this means that the collapse of the perturbation  occurs only if the three axes collapse at the very beginning of the evolution of the perturbation. While this is certainly a sufficient condition, it is clear from Fig. \ref{Fig_Evol_axis} (top left panel) that it is not a necessary condition.

\subsection{Modification of the collapse criterion}
\label{Sec_collapse_criterion}

From the axis evolutions described in the previous section, we can see that the collapsing barrier cannot  be determined by considering a criterion on only one axis. Even if it expands at the beginning of the evolution, the increase in the density of the structure due to the collapse of the  two other axes can cause the third one to collapse. 
To determine the collapse threshold of the structure, we need to understand what happens when the aspect ratio of the ellipsoid becomes extremely large, that is, when only two axes have collapsed by the end of the integration. This corresponds to a very long and  dense filament, which is unphysical. Indeed, such a structure will have fragmented into several substructures long before it reaches this stage. 

\begin{figure}
   \centering
   \includegraphics[width=\columnwidth]{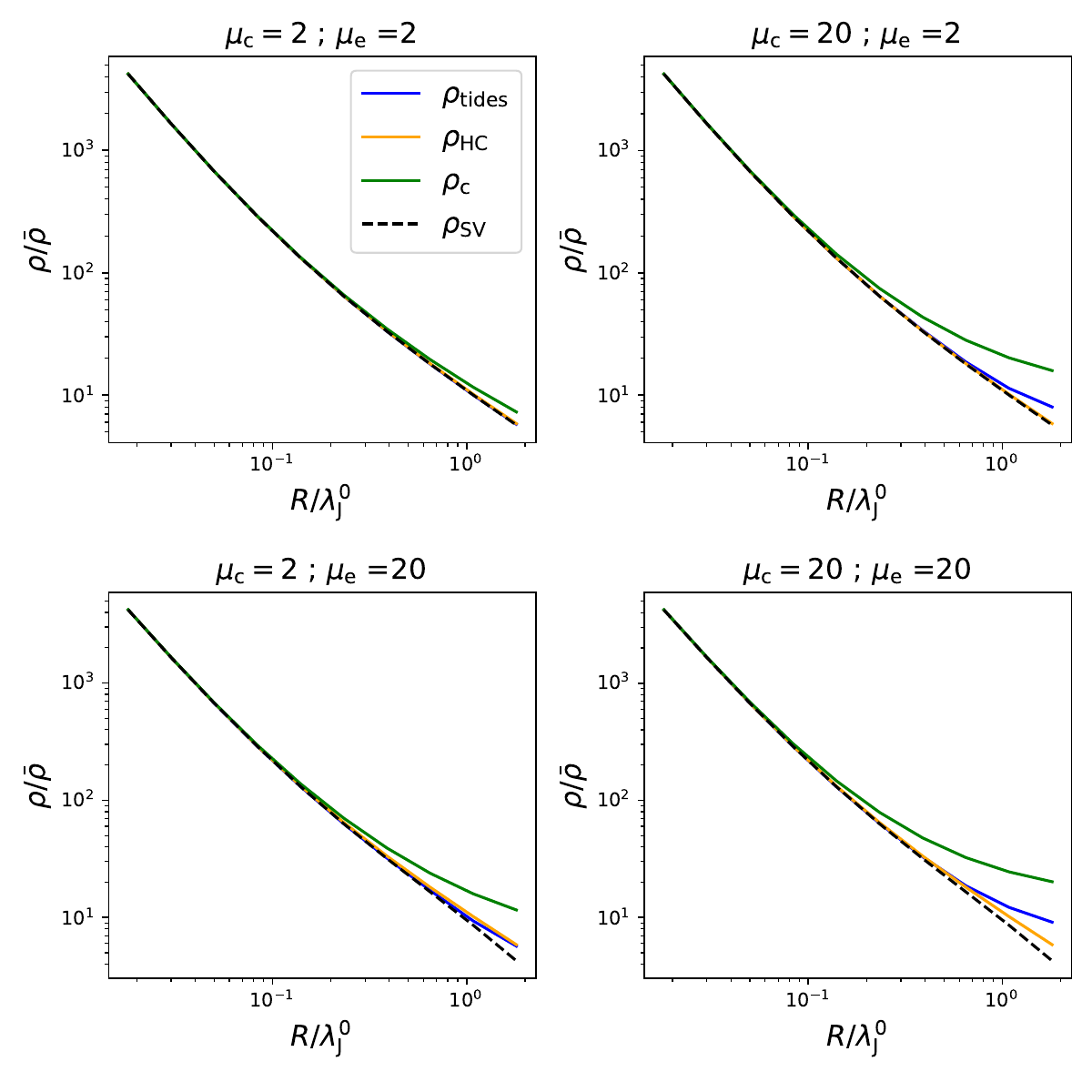}
   \caption{Collapsing barrier (blue line) for different tidal field strengths based on the evolution of the triaxial ellipsoid. This barrier is compared with the barrier computed by \cite{Hennebelle_AnalyticalTheoryInitial2008} without tides (orange line), the critical density for collapse of the \(c\) axis (green line) as in \cite{Colman_OriginPeakStellar2020}, and the global critical density, \(\rho_{\rm SV}\), computed from the scalar virial theorem (dashed black line).}
   \label{Fig_Corr_barr}
\end{figure}

After \cite{Inutsuka_SelfsimilarSolutionsStability1992}, the instability that occurs in an infinite supercritical cylinder is characterized by a most unstable wavelength \(\lambda_{\rm max}\simeq 8\, R_{\rm c}\), where \(R_{\rm c}\) is the inner radius of the isothermal cylinder. The smallest unstable mode is \(\lambda_{\rm crit}\simeq 4\, R_{\rm c}\). Filaments with a line mass greater than \(m_{\rm crit} = 2\sigma_{\rm tot}^2/G\), with \(\sigma_{\rm tot}^2 = c_{\rm s}^2+v_{\rm RMS}^2\), and a much longer aspect ratio will not form because they will fragment. Therefore, we used the following criterion to determine whether or not a structure will lead to the formation of cores: when the aspect ratio $c/a$ or $c/b$ of the structure reaches 8, if the three axes collapse (i.e., \(\partial^2 a_i^2/\partial t^2<0\)), then bound cores will form. This does not mean that the structure will collapse directly to this stage, but as the structure fragments, each fragment with an aspect ratio equal to or larger than \(8\) will collapse along its three axes, as expected from the linear stability analysis. In this situation, multiple bound structures are expected to form from the initial perturbation. To verify that a value of 8 for the critical aspect ratio does not change much the results, we checked that modifying it has a limited impact on the derived collapsing barrier (blue curve in Fig. \ref{Fig_Corr_barr}). Taking a critical aspect ratio of 4 or 16 instead of 8 modifies the barrier by at most 20\%. Our result is thus not significantly impacted by this choice.

In Fig. \ref{Fig_Corr_barr} we show the collapse threshold for a perturbation in a tidal field characterized by the parameters \(\mu_{\rm c}\) and \(\mu_{\rm e}\). We compared this barrier with the one for a spherical structure computed from the scalar virial theorem:
\begin{equation}
   \tilde{\rho}_{\rm SV} = \frac{15}{4\pi}\frac{1+\mathcal{M}_\star^2 \tilde{R}^{2\eta}}{\tilde{R}^2}-\frac{3-\gamma}{4\pi}\mu_{\rm e}.
\end{equation}
Compared to \(\rho_{\rm SV}\), the barrier is indeed increased because of the anisotropy of the tidal field. This confirms that the scalar virial theorem gives a poor estimate of the barrier when anisotropic processes are important. Comparing with the barrier \(\rho_{\rm HC}\) without tides, we see that the modification is very limited, on the order a few tens of percent, even in the strong tidal regime. Therefore, tides do not significantly modify the critical density required for a perturbation to collapse, and thus do not significantly affect { the IMF that would be predicted by models that involve a collapse density threshold such as \cite{Hennebelle_AnalyticalTheoryInitial2008} or \cite{Hopkins_StellarInitialMass2012}}.
Furthermore, it is worth noting that the barrier is well below the one that would be derived if we impose for the global collapse condition of the fluctuations that the three axes must collapse, as suggested by \cite{Colman_OriginPeakStellar2020}. This is because the collapse of two of the three axes increases the density of the structure, making it unstable. 

\section{The dynamical evolution of an ellipsoid with rotational support in a tidal field}
\label{Sec_Tides_env}

The case presented above is the one usually studied in the literature, but it is not physically satisfactory. Indeed, a force is needed to balance  the gravitational acceleration exerted by the central object. Without this force, the perturbation will be accreted by the central core on a dynamical timescale, \(\tau_{\rm dyn}\), which is also the orbital timescale. Then, the question is how does a force that prevents global accretion of the perturbation by a central core affect the stability of the perturbation?

The mechanism that is likely to play a significant role is the orbital rotation of the perturbation around the central object. The presence of the tidal field will affect the perturbation's own motion, which will eventually become tidally synchronized with its orbital motion because of  tidal dissipation. This case is usually studied as the Roche ellipsoid \citep{Chandrasekhar_EllipsoidalFiguresEquilibrium1969, Lai_EllipsoidalFiguresEquilibrium1993}.
The timescale for synchronization depends on the strength of the tides and can be estimated as follows \citep{Gladman_SynchronousLockingTidally1996, Leconte_TidalHeatingSufficient2010}:
\begin{equation}
   \frac{\tau_{\rm syn}}{\tau_{\rm dyn}} \simeq 0.1 \frac{Q}{k_2}\frac{\tau_{\rm dyn}^2}{\tau_{\rm ff}^2},
\end{equation}
where \(k_2\) is the tidal Love number taken to be equal to 3/2 for very deformable structures. \(Q\) is the tidal dissipation parameter estimated to be \citep{Barker_NonlinearEvolutionTidal2013}
\begin{equation}
   Q\simeq \left(\frac{M_{\rm p}}{M_{\rm c+e}}+1\right)\left(\frac{\tau_{\rm dyn}}{\tau_{\rm ff}}\right)^4.
\end{equation}
For a perturbation of mass smaller than the mass of the central core, we finally get 
\begin{equation}
   \frac{\tau_{\rm syn}}{\tau_{\rm dyn}}\simeq 0.1\left(\frac{\tau_{\rm dyn}}{\tau_{\rm ff}}\right)^6,
\end{equation}
In the  two next sections we discuss the two associated regimes, where the perturbation is tidally locked to the central object (strong tidal regime) and where it is not (weak tidal regime). 

\subsection{A perturbation not tidally locked around the  central object}

If the free-fall timescale of the perturbation is small compared to the dynamical timescale (\(\tau_{\rm ff}<\tau_{\rm dyn}\)), corresponding to the weak tidal regime (see Sect. \ref{Sec_Dynamical_evolution}), the perturbation will not be tidally locked. In this case, the perturbation will barely have time to move along its orbit during its evolution and the axis \(c\) will remain oriented toward the central core. This case is equivalent to the one studied in Sect. \ref{Sec_Dynamical_evolution}, restricted to weak tides. In this regime, the tides are not able to significantly stabilize the perturbation against its collapse. As shown in Fig. \ref{Fig_Corr_barr}, the increase in the barrier compared to the one in the absence of tides is only on the order a few percent.

\subsection{A tidally locked perturbation}

If the dynamical timescale is small relative to the collapse timescale of the perturbation (\(\tau_{\rm ff}>\tau_{\rm dyn}\)), corresponding to the strong tidal regime, then the perturbation will synchronize almost immediately. As a consequence, the side of the perturbation facing the central object will always be the same. In this case, an additional support must be taken into account, namely the orbital rotational support. 
Thus, we added to Eq. (\ref{eq_Viral_eq})  the tensor associated with the rotation:
\begin{equation}
   \mathcal{R} = \int \rho\Omega \vec{r}\cdot  (x\vec{e}_x+ z\vec{e}_z){\rm d}V,
\end{equation}
where \(\vec{\Omega} \parallel \vec{e}_y\) is the rotation axis of the perturbation. When the perturbation is tidally locked with its orbit, \(\Omega\) takes the Keplerian value:
\begin{equation}
   \Omega^2 = \frac{G}{r_0^3}(M_{\rm c+e}(r_0)+M_{\rm p}),
\end{equation}
where \(M_{\rm c+e}(r_0)\) is the mass contained within the radius \(r_0\) from the central object.
The tensor reads
\begin{equation}
   \label{eq_Virial_rotation_tensor}
   \mathcal{R} = \frac{M_{\rm p}}{5}\Omega^2\left[\begin{array}{ccc}
      a^2 & 0 & 0 \\
      0 & 0  & 0 \\
      0 & 0 & c^2
      \end{array}\right].
\end{equation}
We deduced the  value of the trace of the tidal plus rotational tensor to determine the stability of a spherical perturbation of radius \(R\):
\begin{equation}
   \text{Tr}(\mathcal{T}+\mathcal{R}) > \frac{G R^2 M_{\rm p}}{5}\left[\frac{2 M_{\rm c}}{r_0^3}+4\pi \rho_{\rm e}(r_0)\right].
\end{equation}
Equality is reached when \(M_{\rm p}\ll M_{\rm c+e}(r_0)\). The calculation performed here is only approximate because we have treated the total mass enclosed in the envelope as a point mass, but it shows that the rotation generated by tidal locking provides sufficient support to eventually disrupt the whole structure. We thus emphasize the fact that both the actions of tides and rotation are required to disrupt a perturbation in the strong tidal regime.  

More quantitatively, we computed the equation of evolution of the triaxial ellipsoid taking into account both the tides and the rotational support. They have the same form as Eq. (\ref{eq_sys_axis_evol}) but without azimuthal symmetry:
   \begin{align}
      \label{eq_sys_axis_evol_rot}
      \frac{1}{2}\frac{\partial^2 \tilde{a}^2}{\partial\tilde{t}^2} &=5(1+\mathcal{M}_\star^2 \tilde{a}^{2\eta})-2\pi\alpha_1 \tilde{c}^2\frac{\rho}{\bar{\rho}} , \nonumber \\
      \frac{1}{2}\frac{\partial^2 \tilde{b}^2}{\partial\tilde{t}^2} &=5(1+\mathcal{M}_\star^2 \tilde{a}^{2\eta})-2\pi\alpha_2 \tilde{c}^2\frac{\rho}{\bar{\rho}}-\tilde{b}^2(\mu_{\rm c}+\mu_{\rm e})  ,  \\
      \frac{1}{2}\frac{\partial^2 \tilde{c}^2}{\partial\tilde{t}^2} &=5(1+\mathcal{M}_\star^2 \tilde{a}^{2\eta})-2\pi\alpha_3 \tilde{c}^2\frac{\rho}{\bar{\rho}}+\tilde{c}^2\left(3\mu_{\rm c}+\gamma\mu_{\rm e}\right). \nonumber
\end{align}
We assumed that the mass of the perturbation is small compared to the mass of the central object inducing the tides. 

\begin{figure}
   \centering
   \includegraphics[width=\columnwidth]{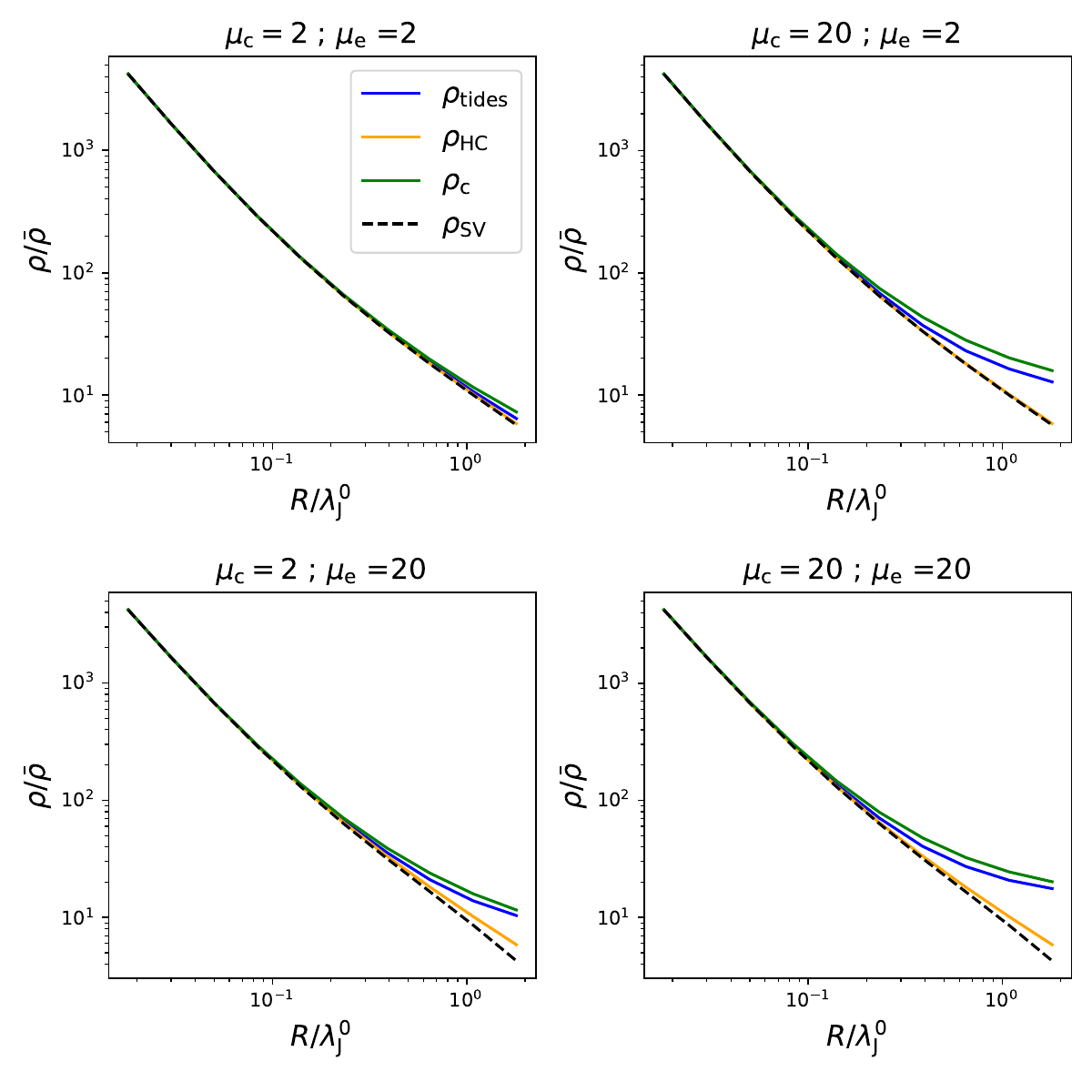}
   \caption{Same as Fig. \ref{Fig_Corr_barr} but for a tidally locked perturbation. In addition to the action of the tides on the dynamics of the perturbation, the rotational support is taken into account.}
   \label{Fig_Corr_barr_rot}
\end{figure}

Using the same criterion  as previously to calculate the barrier from the evolution, we plot the modified collapse density threshold in Fig. \ref{Fig_Corr_barr_rot}. Compared to Fig. \ref{Fig_Corr_barr}, the barrier is increased. Compared to the usual barrier without tides and rotation, these mechanisms can provide a substantial support that can modify the dynamics of an unstable perturbation. We note that the derived barrier is close to \(\rho_c\), even though \(\rho_c\) is calculated without the rotational support. This suggests that the criterion proposed by \cite{Colman_OriginPeakStellar2020}, although not obtained consistently due to the lack of rotation, may be valid in the strong tidal regime, that is, for tidally locked perturbations, because of the rotation. 

\section{Discussion}
\label{Sec_Discussion}

\subsection{Which tidal regime for a perturbation formed close to a Larson core?}

From the previous investigation, we have seen that the effect of tides on the dynamics of a structure depends on their strength. In the weak tidal regime (see above), the dynamics of the fluctuation can be assumed to be non affected by its orbital motion. The tides will barely contribute to the dynamics of the structure and the collapsing barrier will only be increased by a few percent. In the strong tidal regime, the perturbation will be stabilized against collapse because of the double effects of the tides and the rotation induced by  tidal synchronization. In this case, the collapsing density threshold can be considerably larger than in the absence of tides. 

The type of regime, weak or strong, depends on  (i) the mass of the central object,  (ii) the distance between this object and the perturbation, and  (iii) the density of the perturbation. We can examine which regime is relevant in the case of a perturbation formed by turbulence in the vicinity of a { first} Larson core of mass \(M_{\rm L}=0.02 \text{M}_\odot\) according to the scenario examined by \cite{Lee_StellarMassSpectrum2018}. In theories of star formation such as \cite{Hennebelle_AnalyticalTheoryInitial2008} and \cite{Hopkins_StellarInitialMass2012}, a perturbation formed at the scale \(R\) will collapse if its density is larger than a certain threshold given by Eq. (\ref{eq_HC_collapse}).
The weak tidal regime is characterized by
\begin{equation}
   \tilde{\rho}=\frac{\rho}{\bar{\rho}}>\frac{15}{4\pi}\frac{1+\mathcal{M}_\star^2 \tilde{R}^{2\eta}}{\tilde{R}^2}>\frac{M_{\rm L}}{r_0^3\bar{\rho}}+4\pi {\tilde \rho}_{\rm e}(r_0).
\end{equation}
The first inequality is the collapse condition Eq. (\ref{eq_HC_collapse}). The density profile of the envelope is taken to be the one of the singular isothermal sphere and the background conditions for mean density and turbulence are the one given in Sect \ref{Sec_tensorial_virial_th}.
For \(r_0>R\), (i.e., when the density perturbation  lies outside the central core),  the critical mass for an unstable perturbation, { for the typical numerical values given at the end of Sect. \ref{Sec_tensorial_virial_th}}, must thus verify
\begin{equation}
   M_{\rm p}(R)>2.7 M_\odot,
\end{equation}
which corresponds to \(\tilde{R}>0.1\). Therefore, large density perturbations, which lead to the formation of massive stars ($>2.7 M_\odot$), will always be in the weak tidal regime. Such perturbations will never be significantly affected by the presence of a nearby collapsing object. 

\begin{figure}
   \centering
   \includegraphics[width=\columnwidth]{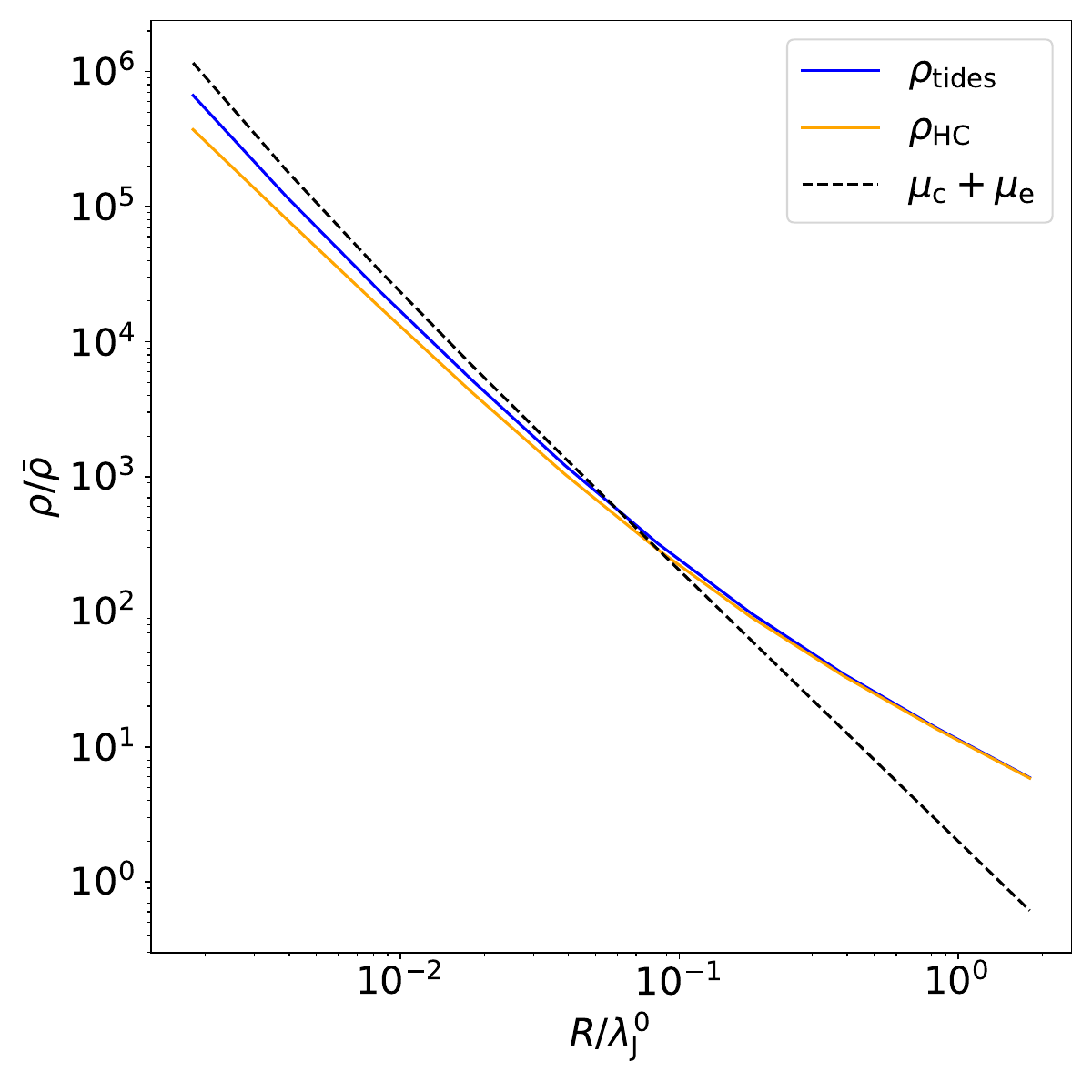}
   \caption{Collapsing barrier (blue) for perturbations at the point closest to the Larson core, compared with the barrier in the absence of tides (Eq. \ref{eq_HC_collapse}; orange). The sum of the two tidal coefficients is shown by the dashed black line. As seen, small perturbations are indeed in the strong tidal regime, which justifies the need to include the rotational support in the calculations.}
   \label{Fig_Corr_barr_Larson}
\end{figure}

In contrast, because of their smaller radius, unstable densities of lower masses can be closer to the central core, and thus be in the strong tidal regime. Their dynamics can be affected by the tides, eventually leading to a tidal radius around the Larson core where structure formation is very unlikely. As examined previously, such a situation would imply a significant increase in the collapse threshold condition for perturbations formed near the Larson core if they are tidally locked. We computed such a threshold condition for a perturbation of radius \(R=r_0\) formed at a distance \(r_0\) from the central core, which is the minimum distance for such a structure. As we are in the strong tidal regime, the effect of rotation is taken into account. The barrier is plotted in Fig. \ref{Fig_Corr_barr_Larson}. We clearly see that the tides hardly modify the usual collapsing barrier (Eq. \ref{eq_HC_collapse}), even in the present case where the perturbation lies just at the boundary of a Larson core. The barrier is increased at most by a factor of $\sim$ { 1.5-1.8} for \(\tilde{R}<10^{-2}\).
Since the perturbation is considered here to be at the boundary of the Larson core, the tidal strength can be underestimated, when using the tidal approximation, if the perturbation is deformed. In Appendix \ref{App_tidal_approx} we quantify such a possible underestimation of the tidal strength. In the very extreme situation where the perturbation is touching the central point mass of the Larson core and reaches the maximum possible deformation (i.e., an aspect ratio of 8; see Sect. \ref{Sec_collapse_criterion}), we find that the barrier is increased by a factor of at most 10 compared to the usual HC08 barrier (see Eq. \ref{eq_HC_collapse}).  We stress, however, that such a case is very unlikely. Indeed, in this case, the perturbation will be accreted on a very short timescale (see Sect. \ref{Sec_typical_mass_CMF}).

We note that for unstable perturbations, the free-fall timescale of the perturbation is at most 1.5 times longer than the dynamical timescale (i.e., orbital timescale \(\tau_{\rm dyn}=\sqrt{\frac{r_0^3}{GM_{\rm c+e}}}\)). Consequently, considering only equilibrium tides is a good assumption. Indeed, in order to affect the dynamics of the structure, dynamical tides would require a much larger number of orbital periods to set in.

To calculate the barrier shown in Fig. \ref{Fig_Corr_barr_Larson}, we used the same fragmentation criterion as the one described in Sect. \ref{Sec_collapse_criterion}. However, when the density of the perturbation is close to the adiabatic density \(n_{\rm ad}\), the proposed scenario becomes dubious. Indeed, the collapse of the ellipsoid is due to the increase in its density following the contraction of the smallest axis under the action of the tides. This contraction is only possible if the density is below the adiabatic density. If this is not the case, the contraction of the smallest axis will stop, leading to a possible disruption of the structure due to the tidal extensive component. This could happen for a perturbation with a mass on the order the first Larson core. For such a perturbation, a collapse criterion based on the fragmentation condition described in Sect. \ref{Sec_collapse_criterion} is irrelevant, since a Larson core cannot fragment. 
In that case, we consider the following collapse criterion: when the density of the perturbation reaches the adiabatic density or an aspect ratio of 4, all axes should collapse (i.e., \(\partial^2 a_i/\partial t^2<0\)). The second condition ensures that no fragmentation occurs (\citealt{Inutsuka_SelfsimilarSolutionsStability1992}; see our Sect. \ref{Sec_collapse_criterion}). With this collapse criterion,{ which is more conservative than the one used in Sect. \ref{Sec_collapse_criterion}}, the derived threshold is the same as the one shown in Fig. \ref{Fig_Corr_barr_Larson}. This indicates that our results are robust.

\subsection{An estimate of the characteristic stellar mass?}
\label{Sec_typical_mass_CMF}

Tides have been proposed as a possible explanation for the observed universality of the peak of the IMF{ \citep{Lee_StellarMassSpectrum2018}}. However, as we have seen, this explanation is unlikely to be relevant, since tides do not alter the dynamics of a newly formed perturbation in a star formation environment. We propose another explanation that could account for this universality. 

Although the suggestion of a radius around a Larson core { in which no new density perturbation can collapse} is certainly relevant, the main phenomenon that determines this radius needs to be clearly identified. In the vicinity of a core, a perturbation will not have time to form if the accretion velocity toward the central object is larger than the velocity of the compressive shock responsible for its formation. In such a case, the induced gas overdensity will be accreted by the central object before it has time to become unstable. This condition can be written as
\begin{equation}
   v_{\rm dyn}(r_0)>v_{\rm RMS}.
\end{equation}
We assumed that the density fluctuation of scale \(R=r_0\) formed at  distance \(r_0\) from the Larson core obeys the Larson velocity relation. Thus, the threshold radius, \(r_0\), verifies
\begin{align}
   \frac{GM_{\rm c+e}}{r_0} =V_0^2\left(\frac{r_0}{1\text{pc}}\right)^{2\eta}
   \implies \tilde{r}_0=\sqrt{\frac{4\pi \tilde{M}_{\rm c}}{9\mathcal{M}_\star^2}},
\end{align}
taking \(\eta=0.5\). The mass contained within the radius \(r_0\) that will be accreted by the central object is given by\begin{equation}
   \tilde{M}_{\rm c} = \tilde{M}_{\rm L} +\frac{3}{2\pi}\tilde{r}_0.
\end{equation}
We finally end up with the following equation:
\begin{equation}
   \tilde{M}_{\rm c} = \tilde{M}_{\rm L} +\sqrt{\frac{\tilde{M}_{\rm c}}{\pi\mathcal{M}_\star^2}} \implies M_{\rm c+e} \simeq \frac{4}{3}\frac{c_{\rm s}^3}{\mathcal{M}_\star^2 \bar{\rho}^{1/2} G^{3/2}}.
\end{equation}

Under the typical initial conditions used in the simulations from \cite{Lee_StellarMassSpectrum2018} and \cite{Colman_OriginPeakStellar2020}, with \(\bar{\rho}=3\times10^5 \text{cm}^{-3} , c_{\rm s} = 190 \text{m/s}\) and \(\mathcal{M}_\star = 1.5\), we find a peak at 0.1 M\(_\odot\).
This result is very close to the observed peak of the sink mass function in those simulations. 

However, under standard star forming conditions considered in that paper (see Sect. \ref{Sec_tensorial_virial_th}), we predict a peak at 1 M\(_\odot\).
This result is close to the observed characteristic mass of the core mass function, on the order 0.6 M\(_\odot\) \citep{Konyves_CensusDenseCores2015}. Looking at the values relevant to massive early type galaxies \citep{Chabrier_VariationsStellarInitial2014}, we find that the predicted mass of the peak of the core mass function is reduced by a factor of 3-5 compared to the one of the Milky Way. This is in qualitative agreement with the observations, which reveals an excess of low-mass stars for these galaxies, and the predictions of the HC theory \citep{Chabrier_VariationsStellarInitial2014}. Unfortunately, observations for these galaxies cannot so far determine the value of the peak of the IMF.

The picture presented here suggests that the perturbations formed in the envelope are generated by turbulence. For this scenario to be realistic, the turbulence should be strong enough at the core scale, that is to say, the sonic scale should be at least smaller than the typical size of the envelope surrounding the Larson core. This condition is the one taken in the simulations discussed above, where the sonic length is on the order \(10^{-3}\) pc. However, all observations of Milky Way star forming regions of the velocity dispersion at the core scale \citep{Auddy_MagneticFieldStructure2019, Redaelli_IdentificationPrestellarCores2021, Choudhury_TransitionCoherentCores2021, Li_ALMASurvey702023} suggest that the turbulence is at most transonic at such small scales. Turbulence, however, is still above what would be expected for a turbulent cascade, probably because other processes come into play, such as adiabatic heating \citep{Robertson_ADIABATICHEATINGCONTRACTING2012, Hennebelle_AmplificationGenerationTurbulence2021}, which maintain the turbulence at this transonic level. Because the turbulence level at such small scales (i.e., below the sonic scale) can only produce a tiny fraction of very dense gas (see \citealt{Federrath_SonicScaleInterstellar2021a}, their Fig. 3), unstable perturbations in the envelope close to the Larson core are unlikely to form. This suggests that the initial condition taken in the two simulations discussed above are not representative of star forming regions.

\subsection{Comparison with other studies}

Based on the scalar virial theorem, \cite{Lee_StellarMassSpectrum2018} show that the tides can stabilize a perturbation formed near the first Larson core. { In their study, the stabilizing effect of the tides is very strong, increasing the barrier by more than one order of magnitude}. This seems to contradict the result presented in Sect.\ref{Sec_Global_energy_budget}, where we show that the scalar virial theorem suggests that the tides favor the collapse of the structure (\(\text{Tr}(\mathcal{T})<0\)). This result holds for any spherically symmetric profile of the perturbation. In \cite{Lee_StellarMassSpectrum2018}, however, the perturbation studied by the author is not spherically symmetric but can be written as\begin{equation}
   \rho_{LH}(\delta r)\propto \frac{1}{r_0^2+\delta r^2}+\frac{\zeta}{r_0^2},
\end{equation}
where \(\zeta\) is a constant that determines the amplitude of the perturbation. As done in their Appendix E, the radial tensor component associated with the tides exerted by the central point mass Larson core can be expressed as\begin{align}
   \mathclap{\mathcal{T}_{rr}} \quad  &= \int_V \rho_0 r_\star^2 \left(\frac{1}{r_0^2+\delta r^2}+\frac{\zeta}{r_0^2}\right)GM_{\rm c}\left[ -\frac{\vec{r}_1}{r_1^{3}} - \frac{\vec{r}_0}{r_0^3} \right]\cdot \delta\vec{r} dV \nonumber \\
   &= -2\pi \rho_0 \left(\frac{r_\star^2}{R^2}\right) R^2 GM_{\rm c} \left[
      2\text{Arctanh}(r_{\rm p}) - \frac{6 r_{\rm p} - 4 r_{\rm p}^3}{3(1-r_{\rm p}^2)} \right.\notag \\
      & \quad \left.+
      \frac{r_{\rm p}(3-r_{\rm p}^2)}{4}+\frac{r_{\rm p}^4+2r_{\rm p}^2-3}{4}\text{Arctanh}(r_{\rm p})\right],
\end{align}
where \(\vec{r}_1 = \vec{r}_0+ \vec{\delta}_r\), \(r_{\rm p}=R/r_0\), \(r_0\) the distance between the center of mass of the perturbation and the one of the central object and \(r_\star\) a constant.
We can verify that \(\mathcal{T}_{rr}\) would be equal to 0 if the structure had a uniform density.
Then, their result is based on this particular form of symmetry breaking. In particular, when \(R\to r_0\),
\begin{equation}
   \mathcal{T}_{rr} \underset{r_{\rm p}\to 1}{\sim} \frac{1}{(1-r_{\rm p})}.
\end{equation}
A perturbation with this profile, characterized by \(R \simeq r_0\), will thus be subject to a very strong global perturbing tidal force. However, this force decreases very rapidly if \(r_{\rm p}\) is small enough (i.e., \(r_{\rm p}<0.5\)):
\begin{equation}
   \mathcal{T}_{rr} \underset{r_{\rm p}\to 0}{\sim} r_{\rm p}^7.
\end{equation}
In order to get a significant stabilizing effect from the tides, the authors must thus put the perturbation very close to their central object, that is, the radius of the perturbation is very close to the distance between the center of the Larson core and the center of mass of the perturbation (i.e., \(R \simeq r_0\); see their Fig. 10.).
In this case, the perturbation is likely to be accreted by the central core. Indeed, as  discussed in Sect. \ref{Sec_typical_mass_CMF}, perturbations very close to the core are unlikely to form due to the large accretion velocity compared to the turbulent velocity.

Several numerical studies and observations describe the tides as a key process that plays an important role in star formation, and suggest  in particular that they are responsible for the universality of the peak of the IMF. Their argument is based on two ingredients: first, the existence of an accretion radius around the first Larson core. Second, the fact that this radius is determined by the tides. The first point is studied in detail by \cite{Hennebelle_HowFirstHydrostatic2019}, who varied the minimum imposed distance between two sink particles. They show that the peak of the sink distribution is shifted toward high masses as the minimum distance between two sinks is increased. This suggests that the characteristic stellar mass is determined by the amount of gas that a Larson core can accrete. 
However, there is no evidence that this accretion radius is set by the tides. To date, and to the best of our knowledge, no study has shown that the collapse of the structures is prevented by the presence of strong tides. To carry out such a study, one would first need to identify overdense structures at the beginning of their collapse and calculate the external tidal field acting on these structures, as done for example by \cite{Ganguly_SILCCZoomDynamicBalance2024}. If correlations are found between the strength of the tides and the properties (namely density) of the structures, this would imply that the tides play a significant role. Conversely, if no correlation is found (i.e., collapsing structures are found despite being exposed to a strong tidal field), this would mean that the tides do not play a significant role.

\section{Conclusion}

In this study, based on the tidal equations derived in \cite{Chabrier_ConsistentExplanationUnusual2024} and using a fragmentation threshold based on \cite{Inutsuka_SelfsimilarSolutionsStability1992}, we discuss in detail the role of tides in the context of star formation, considering them in the context their entire dynamical and anisotropic effects. We study a turbulence-induced density perturbation formed in the envelope of a central core. We determine the effect the tidal field induced by the core has on the evolution of the perturbation. We define two regimes (see Eqs. \ref{eq_def_weak} and \ref{eq_def_strong}), weak and strong tidal regimes, based on the strength of the tidal force relative to the self-gravity of the density perturbation.

We show that this anisotropy significantly limits the effect of the tides: in the weak tidal regime, the increase in the collapse barrier is small, much smaller than the usual prescriptions used in the literature, which generally consider the tidal field as isotropic. 
In the strong tidal regime, however, if the perturbation is orbiting the central core, it will start to rotationally  synchronize with its orbital motion. In this case, the perturbation will experience an additional support that will prevent the collapse and eventually disrupt the structure. The increase in the collapsing barrier can be significant compared to that in the absence of tides
depending on the various properties of the global orbital system. 
The disruption of the fluctuation in this regime is a consequence not only of the tides, but of the combined effect of the tides and the rotational support. 

In the context of star formation, where perturbations can form near a Larson core, we show that unstable perturbations can be in either the weak or the strong tidal regime, depending on their mass. However, even when taking rotational support into account, in the case of a strong tidal effect, the collapse barrier in the immediate vicinity of the core is  very modestly affected compared to that in the absence of tides. Therefore, the concept of the tides setting up a tidal radius around a Larson core within which  it is much more difficult for a density perturbation to become gravitationally unstable than if it is lying outside this radius is not correct. This seems to invalidate the results of some studies that have claimed that the collapse barrier could increase by more than an order of magnitude. This is essentially due to the anisotropic nature of the tidal field. {Statistically, the number of structures that may collapse inside or outside this tidal radius is very similar}. Consequently, the universality of the peak of the IMF is unlikely to be related to tidal processes.

Finally, we propose another explanation for the universality of the peak of the IMF. If the velocity of the shock forming a perturbation is lower than its infall velocity onto the central core, the material will not have time to condense before being accreted. This implies that there is an{ ``accretion radius''} around the first Larson core within which a new perturbation is very unlikely to form. The corresponding accreted mass within this radius is in good agreement with the characteristic mass observed in the core mass function.

\begin{acknowledgements} 
   We thank the anonymous referee for constructive feedback that helped to improve the clarity of this work. We also thank Jeremy Fensch, Armand Leclerc and Elliot Lynch for very helpful discussions that improved the clarity of the manuscript. We are also grateful to Patrick Hennebelle for a careful reading of the manuscript and stimulating discussions.
\end{acknowledgements}

\bibliographystyle{aa}
\bibliography{Bib_tides}

\begin{thebibliography}{41}
\expandafter\ifx\csname natexlab\endcsname\relax\def\natexlab#1{#1}\fi

\bibitem[{Auddy {et~al.}(2019)Auddy, Myers, Basu, Harju, Pineda, \&
  Friesen}]{Auddy_MagneticFieldStructure2019}
Auddy, S., Myers, P.~C., Basu, S., {et~al.} 2019, ApJ, 872, 207

\bibitem[{Barker \& Lithwick(2013)}]{Barker_NonlinearEvolutionTidal2013}
Barker, A.~J. \& Lithwick, Y. 2013, MNRAS, 435, 3614

\bibitem[{Chabrier \& Dumond(2024)}]{Chabrier_ConsistentExplanationUnusual2024}
Chabrier, G. \& Dumond, P. 2024, ApJ, 966, 48

\bibitem[{Chabrier {et~al.}(2014)Chabrier, Hennebelle, \&
  Charlot}]{Chabrier_VariationsStellarInitial2014}
Chabrier, G., Hennebelle, P., \& Charlot, S. 2014, ApJ, 796, 75

\bibitem[{Chandrasekhar(1969)}]{Chandrasekhar_EllipsoidalFiguresEquilibrium1969}
Chandrasekhar, S. 1969, Ellipsoidal Figures of Equilibrium

\bibitem[{Chandrasekhar \&
  Fermi(1953)}]{Chandrasekhar_ProblemsGravitationalStability1953}
Chandrasekhar, S. \& Fermi, E. 1953, ApJ, 118, 116

\bibitem[{Chandrasekhar \&
  Lebovitz(1963)}]{Chandrasekhar_EquilibriumStabilityJeans1963}
Chandrasekhar, S. \& Lebovitz, N.~R. 1963, ApJ, 137, 1172

\bibitem[{Chen {et~al.}(2016)Chen, {Amaro-Seoane}, \&
  Cuadra}]{Chen_STABILITYGASCLOUDS2016a}
Chen, X., {Amaro-Seoane}, P., \& Cuadra, J. 2016, ApJ, 819, 138

\bibitem[{Choudhury {et~al.}(2021)Choudhury, Pineda, Caselli, Offner,
  Rosolowsky, Friesen, Redaelli, {Chac{\'o}n-Tanarro}, Shirley, Punanova, \&
  Kirk}]{Choudhury_TransitionCoherentCores2021}
Choudhury, S., Pineda, J.~E., Caselli, P., {et~al.} 2021, A\&A, 648, A114

\bibitem[{Colman \& Teyssier(2020)}]{Colman_OriginPeakStellar2020}
Colman, T. \& Teyssier, R. 2020, MNRAS, 492, 4727

\bibitem[{Dale {et~al.}(2019)Dale, Kruijssen, \&
  Longmore}]{Dale_DynamicalEvolutionMolecular2019a}
Dale, J.~E., Kruijssen, J. M.~D., \& Longmore, S.~N. 2019, MNRAS, 486, 3307

\bibitem[{Federrath {et~al.}(2021)Federrath, Klessen, Iapichino, \&
  Beattie}]{Federrath_SonicScaleInterstellar2021a}
Federrath, C., Klessen, R.~S., Iapichino, L., \& Beattie, J.~R. 2021, Nature
  Astronomy, 5, 365

\bibitem[{Ganguly {et~al.}(2024)Ganguly, Walch, Clarke, \&
  Seifried}]{Ganguly_SILCCZoomDynamicBalance2024}
Ganguly, S., Walch, S., Clarke, S.~D., \& Seifried, D. 2024, MNRAS, 528, 3630

\bibitem[{Gezari(2021)}]{Gezari_TidalDisruptionEvents2021}
Gezari, S. 2021, Annu. Rev. Astron. Astrophys., 59, 21

\bibitem[{Gladman {et~al.}(1996)Gladman, Quinn, Nicholson, \&
  Rand}]{Gladman_SynchronousLockingTidally1996}
Gladman, B., Quinn, D., Nicholson, P., \& Rand, R. 1996, Icarus, 122, 166

\bibitem[{Hennebelle(2021)}]{Hennebelle_AmplificationGenerationTurbulence2021}
Hennebelle, P. 2021, A\&A, 655, A3

\bibitem[{Hennebelle \&
  Chabrier(2008)}]{Hennebelle_AnalyticalTheoryInitial2008}
Hennebelle, P. \& Chabrier, G. 2008, ApJ, 684, 395

\bibitem[{Hennebelle \&
  Falgarone(2012)}]{Hennebelle_TurbulentMolecularClouds2012}
Hennebelle, P. \& Falgarone, E. 2012, A\&AR, 20, 55

\bibitem[{Hennebelle {et~al.}(2019)Hennebelle, Lee, \&
  Chabrier}]{Hennebelle_HowFirstHydrostatic2019}
Hennebelle, P., Lee, Y.-N., \& Chabrier, G. 2019, ApJ, 883, 140

\bibitem[{Hopkins(2012)}]{Hopkins_StellarInitialMass2012}
Hopkins, P.~F. 2012, MNRAS, 423, 2037

\bibitem[{Inutsuka \&
  Miyama(1992)}]{Inutsuka_SelfsimilarSolutionsStability1992}
Inutsuka, S.-I. \& Miyama, S.~M. 1992, ApJ, 388, 392

\bibitem[{Jog(2013)}]{Jog_JeansInstabilityCriterion2013}
Jog, C.~J. 2013, MNRAS, 434, L56

\bibitem[{K{\"o}nyves {et~al.}(2015)K{\"o}nyves, Andr{\'e}, Men'shchikov,
  Palmeirim, Arzoumanian, Schneider, Roy, Didelon, Maury, Shimajiri,
  Di~Francesco, Bontemps, Peretto, Benedettini, Bernard, Elia, Griffin, Hill,
  Kirk, Ladjelate, Marsh, Martin, Motte, Nguy{\^e}n~Luong, Pezzuto, Roussel,
  Rygl, Sadavoy, Schisano, Spinoglio, {Ward-Thompson}, \&
  White}]{Konyves_CensusDenseCores2015}
K{\"o}nyves, V., Andr{\'e}, {\relax Ph}., Men'shchikov, A., {et~al.} 2015,
  A\&A, 584, A91

\bibitem[{Kruijssen {et~al.}(2019)Kruijssen, Dale, Longmore, Walker, Henshaw,
  Jeffreson, Petkova, Ginsburg, Barnes, Battersby, Immer, Jackson, Keto,
  Krieger, Mills, {S{\'a}nchez-Monge}, Schmiedeke, Suri, \&
  Zhang}]{Kruijssen_DynamicalEvolutionMolecular2019}
Kruijssen, J. M.~D., Dale, J.~E., Longmore, S.~N., {et~al.} 2019, MNARS, 484,
  5734

\bibitem[{Lai {et~al.}(1993)Lai, Rasio, \&
  Shapiro}]{Lai_EllipsoidalFiguresEquilibrium1993}
Lai, D., Rasio, F.~A., \& Shapiro, S.~L. 1993, ApJS, 88, 205

\bibitem[{Larson(1969)}]{Larson_NumericalCalculationsDynamics1969}
Larson, R.~B. 1969, MNRAS, 145, 271

\bibitem[{Lebovitz(1961)}]{Lebovitz_VirialTensorIts1961}
Lebovitz, N.~R. 1961, ApJ, 134, 500

\bibitem[{Leconte {et~al.}(2010)Leconte, Chabrier, Baraffe, \&
  Levrard}]{Leconte_TidalHeatingSufficient2010}
Leconte, J., Chabrier, G., Baraffe, I., \& Levrard, B. 2010, A\&A, 516, A64

\bibitem[{Lee \& Hennebelle(2018)}]{Lee_StellarMassSpectrum2018}
Lee, Y.-N. \& Hennebelle, P. 2018, A\&A, 611, A89

\bibitem[{Lequeux {et~al.}(2005)Lequeux, Falgarone, \&
  Ryter}]{Lequeux_InterstellarMedium2005}
Lequeux, J., Falgarone, E., \& Ryter, C. 2005, The Interstellar Medium,
  Astronomy and Astrophysics Library No. 0941-7834 (Berlin ; New York:
  Springer)

\bibitem[{Li(2023)}]{Li_TidesCloudsControl2023}
Li, G.-X. 2023, MNRAS, 528, L52

\bibitem[{Li(2024)}]{Li_ModificationJeansCriterion2024}
Li, G.-X. 2024, MNRAS, 532, 1126

\bibitem[{Li {et~al.}(2023)Li, Sanhueza, Zhang, Guido, Sabatini, Morii, Lu,
  Tafoya, Nakamura, Izumi, Tatematsu, \& Li}]{Li_ALMASurvey702023}
Li, S., Sanhueza, P., Zhang, Q., {et~al.} 2023, ApJ, 949, 109

\bibitem[{Pineda {et~al.}(2022)Pineda, Arzoumanian, Andr{\'e}, Friesen,
  Zavagno, Clarke, Inoue, Chen, Lee, Soler, \&
  Kuffmeier}]{Pineda_BubblesFilamentsCores2022}
Pineda, J.~E., Arzoumanian, D., Andr{\'e}, P., {et~al.} 2022, From {{Bubbles}}
  and {{Filaments}} to {{Cores}} and {{Disks}}: {{Gas Gathering}} and
  {{Growth}} of {{Structure Leading}} to the {{Formation}} of {{Stellar
  Systems}}

\bibitem[{Redaelli {et~al.}(2021)Redaelli, Bovino, Giannetti, Sabatini,
  Caselli, Wyrowski, Schleicher, \&
  Colombo}]{Redaelli_IdentificationPrestellarCores2021}
Redaelli, E., Bovino, S., Giannetti, A., {et~al.} 2021, A\&A, 650, A202

\bibitem[{Robertson \&
  Goldreich(2012)}]{Robertson_ADIABATICHEATINGCONTRACTING2012}
Robertson, B. \& Goldreich, P. 2012, ApJ, 750, L31

\bibitem[{Roy {et~al.}(2014)Roy, Andr{\'e}, Palmeirim, Attard, K{\"o}nyves,
  Schneider, Peretto, Men'shchikov, {Ward-Thompson}, Kirk, Griffin, Marsh,
  Abergel, Arzoumanian, Benedettini, Hill, Motte, Nguyen~Luong, Pezzuto,
  {Rivera-Ingraham}, Roussel, Rygl, Spinoglio, Stamatellos, \&
  White}]{Roy_ReconstructingDensityTemperature2014}
Roy, A., Andr{\'e}, {\relax Ph}., Palmeirim, P., {et~al.} 2014, A\&A, 562, A138

\bibitem[{Tatematsu \& Fujimoto(1990)}]{Tatematsu_DynamicsRotatingGaseous1990}
Tatematsu, Y. \& Fujimoto, M. 1990, Publications of the Astronomical Society of
  Japan, 42, 217

\bibitem[{Usami \& Fujimoto(1997)}]{Usami_TidalEffectsRotating1997}
Usami, M. \& Fujimoto, M. 1997, ApJ, 487, 489

\bibitem[{{Zavala-Molina} {et~al.}(2023){Zavala-Molina}, {Ballesteros-Paredes},
  Gazol, \& Palau}]{Zavala-Molina_EffectTidalForces2023a}
{Zavala-Molina}, R., {Ballesteros-Paredes}, J., Gazol, A., \& Palau, A. 2023,
  MNRAS, 524, 4614

\bibitem[{Zhou {et~al.}(2024)Zhou, Dib, Juvela, Sanhueza, Wyrowski, Liu, \&
  Menten}]{Zhou_GasInflowsCloud2024}
Zhou, J.~W., Dib, S., Juvela, M., {et~al.} 2024, A\&A, 686, A146

\end{thebibliography}

\begin{appendix}

\section{Validity of the tidal tensor approximation}
\label{App_tidal_approx}

The analysis performed in this work is based on the tidal tensor approximation (Eq. \ref{eq_tidal_field_tensor}). This approximation is valid in the limit \(\delta r\ll r_0\). In this appendix we determine more precisely the domain of validity of this approximation. Without any approximation, the tidal force is given by (Eq.\ref{eq_tidal_field_exact})\begin{equation}
   \vec{g}_{\rm T}(\vec{r}_1) = \vec{g}(\vec{r}_1)-\vec{g}(\vec{r}_0).
\end{equation}
The virial tidal tensor is given by
\begin{equation}
   \mathcal{T}_{ij}=\int_{V} \rho(\delta\vec{r}) \vec{g}_{\rm T}^i(\vec{r}_1)\delta r_j {\rm d} V.
\end{equation}
We consider first the role of the central object:
\begin{equation}
   \vec{g}_{\rm T}(\vec{r}_1) = -\frac{G M_{\rm c}}{r_1^2}\frac{\vec{r}_1}{|r_1|}+\frac{G M_{\rm c}}{r_0^2}\vec{e}_z,
\end{equation}
and \(\vec{r}_1 = r_0\vec{e}_z+\delta\vec{r}\) with \(\delta\vec{r} = \epsilon(a\sin(\theta)\cos(\phi)\vec{e}_x + b\sin(\theta)\sin(\phi)\vec{e}_y + c\cos(\theta)\vec{e}_z)\), where \(a, b\) and \(c\) are the three axes of the ellipsoid and \(\epsilon\in[0, 1]\).
As before, we considered a homogeneous ellipsoid and computed the component \(\mathcal{T}_{zz}\): 
\begin{multline}
   \mathcal{T}_{zz} = 2\pi\rho_0 G M_{\rm c}\times \\ \int_V \left(-\frac{r_0+c\epsilon\cos(\theta)}{r_1^3}+\frac{1}{r_0^2}\right)  c\sin(\theta)\cos(\theta)\epsilon^3 abc {\rm d}\theta {\rm d}\epsilon \\
   =-\frac{3}{2}GM_{\rm c}\frac{M_{\rm p}}{c^3} r_0^2 \int_{u=0}^{\frac{c}{r_0}}\int_{x=-1}^{1}\frac{(1+ux)x u^3}{\left(1+u^2\left(\frac{1-x^2}{\lambda^2}+x^2\right)+2ux\right)^{3/2}} {\rm d}x {\rm d}u,
\end{multline}
where we performed the variable change \(u=c\epsilon/r_0\) and \(x = \cos(\theta)\), and we introduced the aspect ratio \(\lambda = c/a\). On the other hand, the component \(\mathcal{T}_{zz}\) obtained from the tensor approximation is (Eq. \ref{eq_Virial_tidal_tensor})\begin{equation}
   \mathcal{T}_{zz}^{\rm approx} = \frac{2}{5}\frac{GM_{\rm c}}{r_0^3}M_{\rm p} c^2.
\end{equation}

\noindent We performed the same calculation for the compressive component \(\mathcal{T}_{xx}\), and we get
\begin{multline}
   \mathcal{T}_{xx}= \\
   -\frac{3}{4}GM_{\rm c}\frac{M_{\rm p}}{c^3} \frac{r_0^2}{\lambda^2} \int_{u=0}^{\frac{c}{r_0}}\int_{x=-1}^{1}\frac{u(1-x^2)}{\left(1+u^2\left(\frac{1-x^2}{\lambda^2}+x^2\right)+2ux\right)^{3/2}} u^3 {\rm d}x {\rm d}u,
\end{multline}
and 
\begin{equation}
   \mathcal{T}_{xx}^{\rm approx} = -\frac{1}{5}\frac{GM_{\rm c}}{r_0^3}M_{\rm p} \frac{c^2}{\lambda^2}.
\end{equation}

\begin{figure}
   \centering
   \includegraphics[scale=0.33]{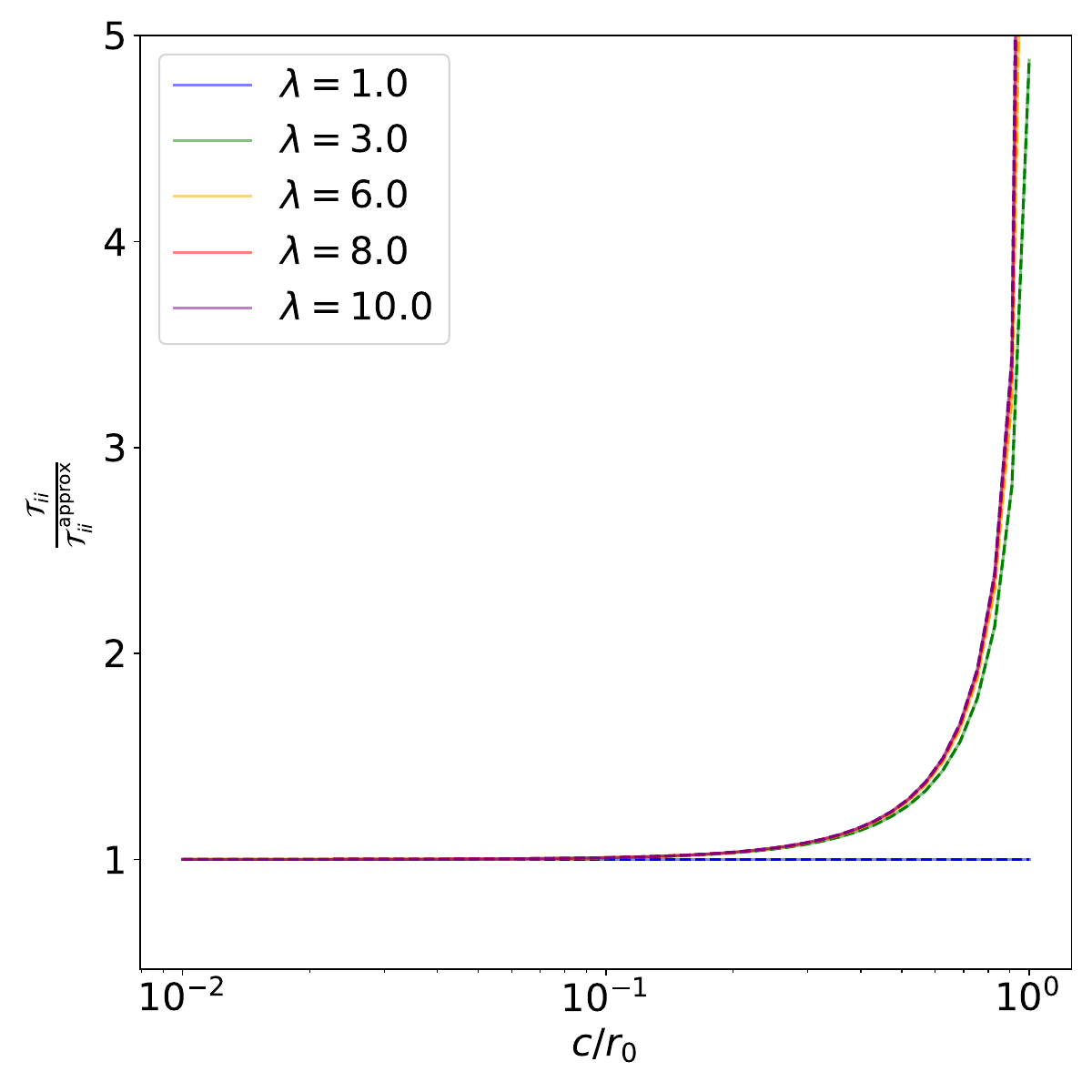}
   \caption{Ratio of the exact virial tidal tensor components \(\mathcal{T}_{zz}\) (solid) and \(\mathcal{T}_{xx}\) (dash) over the approximate ones for various aspect ratios. Unless the radius of the perturbation is similar to the distance between the central core and the center of mass of the perturbation, the approximation is accurate within a factor of 1.3 (for \(c/r_0<0.5\)). The corrections on the longitudinal and orthogonal axes are identical.}
   \label{Fig_Approx_check}
\end{figure}

In Fig. \ref{Fig_Approx_check} we have plotted \(\mathcal{T}_{zz}/\mathcal{T}_{zz}^{\rm approx}\) and \(\mathcal{T}_{xx}/\mathcal{T}_{xx}^{\rm approx}\) for different aspect ratios. The correction only becomes significant when \(c/r_0\to1\) (i.e., when the object is very close to the boundary of the ellipsoidal perturbation). This case is not physical: indeed, we can expect such a perturbation to be accreted very rapidly close to the Larson core. For more reasonable cases, when \(c/r_0<0.5\), the correction is always very small for all aspect ratios, at most on the order a factor of 1.3. This justifies the use of the tensor approximation for the present study. We note that when the aspect ratio is equal to \(\lambda=1\), there is no correction at all, whatever the ratio \(c/r_0\). We can indeed show that
\begin{equation}
   \int_{u=0}^{\frac{c}{r_0}}\int_{x=-1}^{1}\frac{1+ux}{\left(1+u^2+2ux\right)^{3/2}}x u^3 {\rm d}x {\rm d}u = -\frac{4c^5}{15r_0^5},
\end{equation}
which implies that
\begin{equation}
   \mathcal{T}_{zz}=\frac{2}{5}\frac{GM_{\rm c}}{r_0^3}M_{\rm p} c^2
\end{equation}
for a homogeneous sphere. The approximation is thus rigorously valid for a homogeneous sphere. Actually, it is possible to show the stronger result that for any spherically symmetric profile, the two expressions are rigorously equal. Indeed, for the \(T_{zz}\) component, we have
\begin{align}
   \mathcal{T}_{zz} &= -2\pi G M_{\rm c}\int_V \rho(\delta r) \frac{r_0+c\delta r\cos(\theta)}{r_1^3}\sin(\theta)\cos(\theta)\delta r^3 {\rm d}\theta {\rm d}\delta r& \\
   &= -2\pi GM_{\rm c} \int \rho(u)\frac{1+ux}{(1+u^2+2ux)^{3/2}}u^3 x {\rm d}u {\rm d}x \\
   &= 2\pi GM_{\rm c} \int_{0}^{\frac{R}{r_0}} \rho(u)\frac{4}{3}u^4 {\rm d}u.
\end{align}

On the other side, using the tidal tensor approximation:
\begin{align}
   \mathcal{T}_{zz} &= \int \rho(\delta r) \frac{2 G M_{\rm c}}{r_0^3} \delta r^2 \cos(\theta)^2 d\delta r \sin(\theta) {\rm d}\theta {\rm d}\phi \\
   &= 2\pi GM_{\rm c} \int_{0}^{\frac{R}{r_0}} \rho(u)\frac{4}{3}u^4 {\rm d}u.
\end{align}

The same computation can be done for the other components. 
The tidal approximation is this exact for any spherically symmetric perturbation. 

\end{appendix}

\end{document}